\documentclass{nature}

\usepackage{subfigure}
\usepackage{color}
\usepackage[pdftex]{graphicx}
\usepackage[english]{babel}  
\usepackage{amssymb}         
\usepackage[fleqn]{amsmath}         
\usepackage{enumerate}
\usepackage[table]{xcolor}
\usepackage[T1]{fontenc}
\usepackage{caption}
\usepackage{multirow}
\usepackage{makeidx}
\usepackage[normalem]{ulem}
\usepackage{bm}
\usepackage{hyperref}   
\usepackage{booktabs} 
\usepackage{tabularx} 
\usepackage{wrapfig}
\usepackage{float}
\usepackage{geometry}
\usepackage[T1]{fontenc}
\usepackage[table]{xcolor}
\usepackage{makeidx}
\usepackage[normalem]{ulem}

\title{Human Perception of Performance}

\author{Luca Pappalardo$^{1,2}$, Paolo Cintia$^2$, Dino Pedreschi$^1$, Fosca Giannotti$^2$, Albert-L\'aszl\'o Barab\'asi$^{3, 4, 5}$}

\begin{document}

\maketitle

\begin{affiliations}
 \item Department of Computer Science, University of Pisa, Italy
 \item Institute of Information Science and Technologies (ISTI), CNR, Pisa, Italy
 \item CCNR and Physics Department, Northeastern University, Boston, Massachusetts, USA 
 \item Department of Medicine, Harvard Medical School, Boston, Massachusetts, USA
 \item Center of Network Science, Central European University, Budapest, Hungary
\end{affiliations}

\begin{abstract}
Humans are routinely asked to evaluate the performance of other individuals, separating success from failure and affecting outcomes from science to education and sports. Yet, in many contexts, the metrics driving the human evaluation process remain unclear. Here we analyse a massive dataset capturing players' evaluations by human judges to explore human perception of performance in soccer, the world's most popular sport. We use machine learning to design an artificial judge which accurately reproduces human evaluation, allowing us to demonstrate how human observers are biased towards diverse contextual features. By investigating the structure of the artificial judge, we uncover the aspects of the players' behavior which attract the attention of human judges, demonstrating that human evaluation is based on a noticeability heuristic where only feature values far from the norm are considered to rate an individual's performance.
\end{abstract}

Scoring the performance of a chef, a singer, a scientist, an athlete, has always been popular and attractive, determining the performer's chances of success \cite{taylor2012effect, bruin2005save, harackiweicz1989performance, balietti2016peer,vaz2012forecasting, newman2004who, sinatra2016quantifying}. Yet, little is known about the aspects that determine human perception of performance. How do expert reviewers, as well as ordinary people, arrive to their evaluations? To what extent these evaluations are based on objective performance features? How are they affected by subjective biases or contextual influences?

Shedding light on the cognitive process behind human evaluation of performance is an intriguing challenge, attracting scholars from multiple disciplines \cite{morrow2015measurement, ishi1985model, yammarino1997managers, cao2017statistically, tversky1974judgement, tomkins2017reviewer}. Indeed, evaluation plays a fundamental role in performance monitoring, enabling achievement confirmation, training assessment and personal motivation \cite{park2014evaluation, balietti2016peer,clauset2015safe}. By gaining insights into the criteria behind human perception of performance we get a better understanding of what attracts the attention of enthusiasts, practitioners and insiders. This is particularly valuable in sports, where performance scoring highly influences the popularity and the success of clubs, athletes and competitions \cite{franck2008mechanisms, torgler2007shapes, lucifora2003superstar}, with wide economic consequences in advertising revenues and associated markets \cite{herm2014crowd, franck2008mechanisms}. 

Data science offers a novel paradigm to investigate how humans evaluate performance, that we summarize with three words: observe, predict, explain. First, we can collect detailed data via sophisticated sensing technologies which allow us to objectively measure and quantify different aspects of human performance \cite{mello2017bigdata}. 
We can therefore construct features associated with a specific performance and add the outcome of the performance's evaluation assigned by a human expert who witnessed the performance.
Second, we can leverage these observational data linked to human evaluations to construct a predictive model, using machine learning techniques. These models can learn from the observed examples a function that assigns an evaluation outcome to any performance, mimicking the evaluation of the human expert recorded in the data. Depending on the quality of the observed examples and the adequacy of the machine learning model, such a function may accurately predict what would be the expert's evaluation for a given performance, thus providing a high-fidelity artificial intelligence proxy of human expertise.
Third, we may try to explain the obtained machine learning model to discover the features that most influence the evaluation outcome and highlight the rules adopted by the model to score a performance, thus reproducing the logic of the human evaluator.

In this paper we apply this approach to sports performance evaluation, focusing on soccer, the world's most popular sport \cite{dobson2011}. We choose this domain for two reasons: the availability of a wealth of performance and evaluation data \cite{gudmundsson2017spatio, stein2017how}, and the difficulty of evaluating a player's performance, especially when a single judge is tasked to evaluate the performance of all the players in a game. Although data-driven techniques for performance evaluation are increasingly proposed in literature \cite{cintia2015harsh, brooks2016developing, mchale2012development, peel2015predicting, merritt2014scoring,heuer2010soccer}, soccer practitioners, fans and players continue to pay attention to the soccer player ratings, representing human expert-based numeric evaluations of a player's performance compiled by sports newspapers after each game \cite{borden2016soccer}.
Notwithstanding, the mechanisms behind the newspapers' evaluation process and the aspects of performance which most influence the evaluators remain unclear. We therefore collected information about soccer player ratings assigned to every player of a game by reporters from the three most prominent Italian sports newspapers. We then associate each rating to a high-dimensional vector of features extracted by massive data describing any quantifiable aspects of soccer games \cite{gudmundsson2017spatio,stein2017how,cintia2015harsh,cintia2015network}. We use the data to train an artificial judge which learns the relation between technical performance and soccer ratings, hence approximating the human evaluation process. We show that technical features alone -- such as the quality of passes, the number of goals or the dangerousness of a player -- cannot fully explain the human evaluation process. Instead, the inclusion of contextual information of diverse nature -- such as a game's expected result as estimated by bookmakers -- allows the artificial judge to produce a high statistical agreement with human ratings. By inspecting the structure of the artificial judge we find that human evaluation criteria follow a simplistic cognitive process based on a \emph{noticeability heuristic}:
judges first select a limited number of features which attract their attention and then rate a performance based on the presence of noticeable values, i.e., features values far from the norm that can be easily brought to mind. We show that an artificial judge trained on a low-dimensional and noticeability-based representation of performance produces a good statistical agreement with human ratings, revealing the simplicity of the human rating process.

\section*{Results}
\subsection{Data sets and measures (Observe).}
Our dataset consists of soccer logs \cite{gudmundsson2017spatio,stein2017how,cintia2015harsh,pappalardo2017quantifying} describing 760 games of the Italian Serie A in seasons 2015/2016 and 2016/2017. Every game is described by a sequence of events on the field with a total of one million events. Each event -- passes, shots, tackles, dribbles, clearances, goalkeeping actions, fouls, intercepts, aerial duels (see Supplementary Note 1) -- consists of a timestamp, the players involved in the event, the position and the outcome (successful or failed). From these events we derive a large number of features by counting the number of events of a specific type, aggregating events, or combining them through specific functions (Supplementary Note 2) which capture the technical performance of a player $u$ during a game $g$, defined as a $n$-dimensional feature vector: $$\vec{p}(u, g) = [x_1(u, g), x_2(u, g), \dots, x_n(u, g)],$$ where $x_i(u, g)$ is a feature describing a specific aspect of $u$'s technical performance in game $g$. We extract 150 technical performance features, such as the total number of shots by a player in a game (volume of play), the number of successful passes to a teammate (quality of play), or a player's dangerousness during a game (see Supplementary Note 2). Since the technical features have different range of values, we standardize them by computing their z-scores. In total, we analyze $\approx$20,000 technical performance vectors, each describing the performance of a player in a specific game. 

\begin{figure}\centering
\includegraphics[scale=0.75]{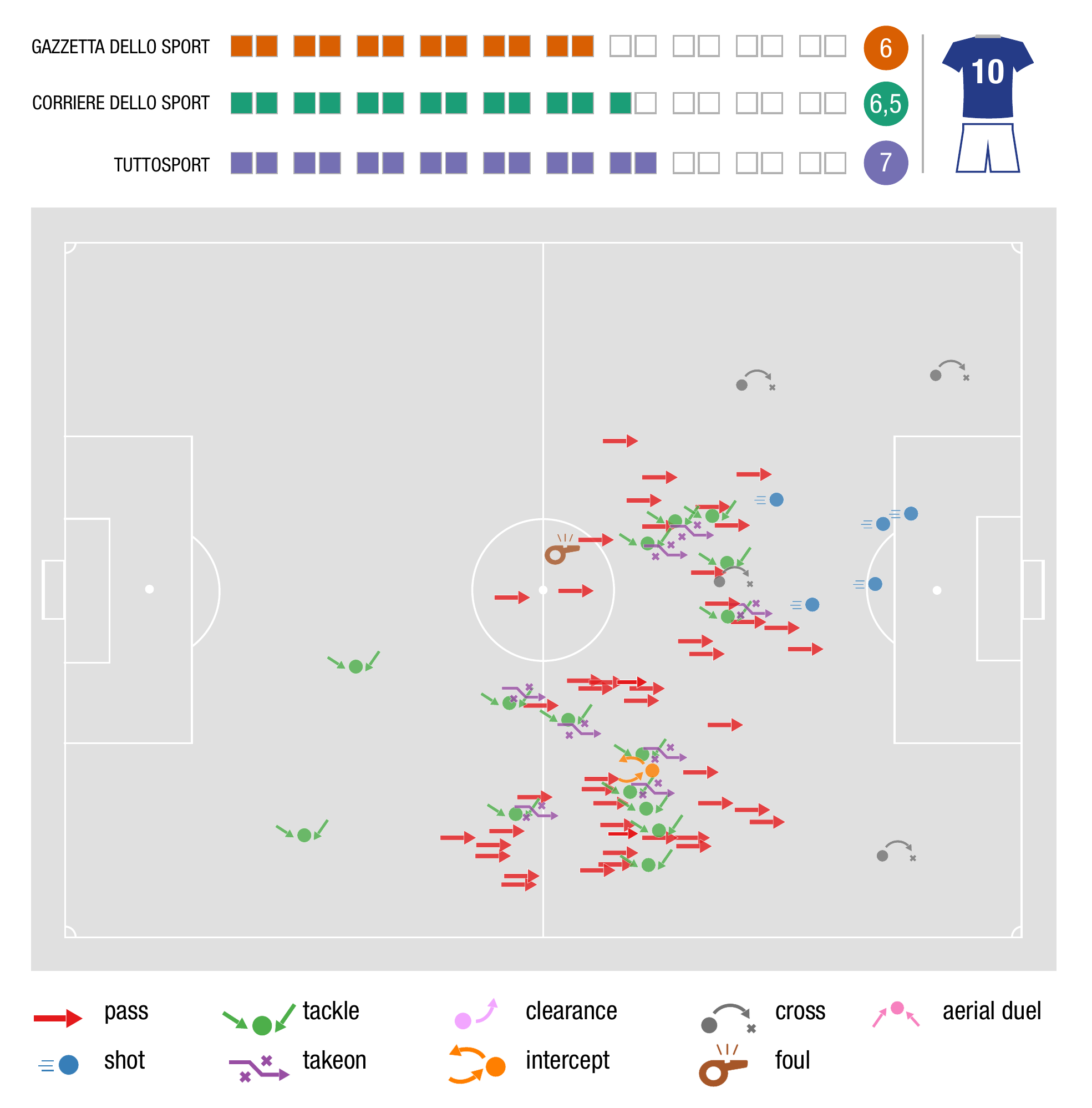}
\caption{\scriptsize \textbf{Technical performance and soccer player ratings.} The image illustrates a portion of the events produced by a player during a game. The game is watched by three sports newspapers' reporters (Gazzetta dello Sport, Corriere dello Sport, Tuttosport) who assign a rating according to their personal interpretation of a player's performance.}
\label{fig:game_visualization}
\end{figure}

Our second data source consists of all the ratings given by three Italian sports newspapers -- Gazzetta dello Sport ($G$), Corriere dello Sport ($C$) and Tuttosport ($T$) -- to each player after each game of the season. A rating $v(u, g, x)$ is assigned to a player $u$ by an anonymous judge (typically a sports reporter) who watched game $g$ for newspaper $x \in \{G, C, T\}$ (see Figure \ref{fig:game_visualization}). These ratings range from 0 (unforgettably bad) to 10 (unforgettably amazing) in units of 0.5 and indicate a judge's perception of a player's performance during a game. 
It is worth noting that both variables and the criteria behind these decision records are unknown, though they hopefully rely on some quantitative aspects of technical performance. 

Though the three newspapers rate the players independently, we find that their rating distributions are statistically indistinguishable (Figure \ref{fig:ratings}a). They are peaked at 6, indicating that most performances are rated as sufficient (see Figure \ref{fig:ratings}b). The soccer rating system is borrowed from the Italian school rating system, where 6 indicates sufficiency reached by a student in a given discipline.
Only $\approx 3\%$ of ratings are lower than 5 and only $\approx 2\%$ of ratings are higher than 7. These outliers refer to rare events of historical importance or deplorable actions by players, such as a 10 assigned to Gonzalo Higua\'in when he became the best seasonal top scorer ever in the history of Serie A, or a 3 assigned to a goalkeeper who intentionally nudged an opponent in the face. In line with the high school rating system, we categorize the ratings in bad ratings (< $6$), sufficient ratings ($\in \{6, 6.5\}$), and good ratings $(> 6.5)$, finding that bad ratings are around three times more frequent than good ratings. 

Ideally, the same performance should result in the same rating for all three judges. In reality, ratings are a personal interpretation of performance, prompting us to assess the degree of agreement between two judges' ratings. We observe a strong Pearson correlation ($r=0.76$) between all newspapers pairs (Figure \ref{fig:ratings}c), though in rare cases two ratings may differ by up to 6 rating units. In Figure \ref{fig:ratings}c we highlight areas of disagreement, i.e., situations where a judge perceives a performance as good while the other perceives it as bad, representing around 20\% of the cases. Also, the Root Mean Squared Error (RMSE) of a rating set with respect to another is $RMSE = 0.50$, meaning that typically the error between two paired ratings is within one unit (i.e., $0.5$) and that there is no systematic discrepancy between the newspapers' ratings (Supplementary Note 3). 
In summary, though situations of disagreement exist, overall we observe a good agreement on paired ratings between the newspapers, finding that the ratings (i) have identical distributions; (ii) are strongly correlated to each other; and (iii) typically differ of one rating unit ($0.5$). 

The key question we wish to address is the consistency of ratings with respect to technical performance, i.e., to what extent similar technical performances are associated with similar ratings \cite{hogson2008examination, mantonakis2009order}. For every pair of technical performances $\vec{p}_i$ and $\vec{p}_j$ ($i \neq j$) we compute their Minkowski distance $d_T(\vec{p}_i, \vec{p}_j)$ (see Supplementary Note 4) and the absolute difference of the corresponding ratings $d_R(\vec{p}_i, \vec{p}_j) = |v_i - v_j|$. We then investigate the relation between $d_T(\vec{p}_i, \vec{p}_j)$ and $d_R(\vec{p}_i, \vec{p}_j)$, finding that the more similar two performances are, the closer are the associated ratings (Figure \ref{fig:ratings}d). This is good news, confirming the existence of a relationship between technical performance and soccer ratings, i.e., judges associate close ratings to performances that are close in the feature space. 

\subsection{Simulating human perception of performance (Predict).}
The criteria behind the rating process of the judges are unknown, since their decisions are a personal interpretation of a player's performance during a game. Here we use machine learning to generate an ``artificial judge'', i.e., a classifier which simulates the human rating process. The artificial judge serves two purposes: \emph{(i)} helps us understand to what extent human decisions are related to measurable aspects of a player's technical performance; \emph{(ii)} helps us build an interpretable model to unveil the rating criteria, i.e., the reasoning behind the judges' rating decisions. Formally, we assume the existence of a function $\mathcal{F}$ representing a rating process which assigns a set of ratings $V$ based on a set of technical performances $P$, i.e., $V = \mathcal{F}(P)$. We use machine learning to infer a function $f$ (i.e., the artificial judge) which approximates the unknown function $\mathcal{F}$, representing the human judge's rating process. 

\begin{figure}
    \centering
    \includegraphics[scale=0.35]{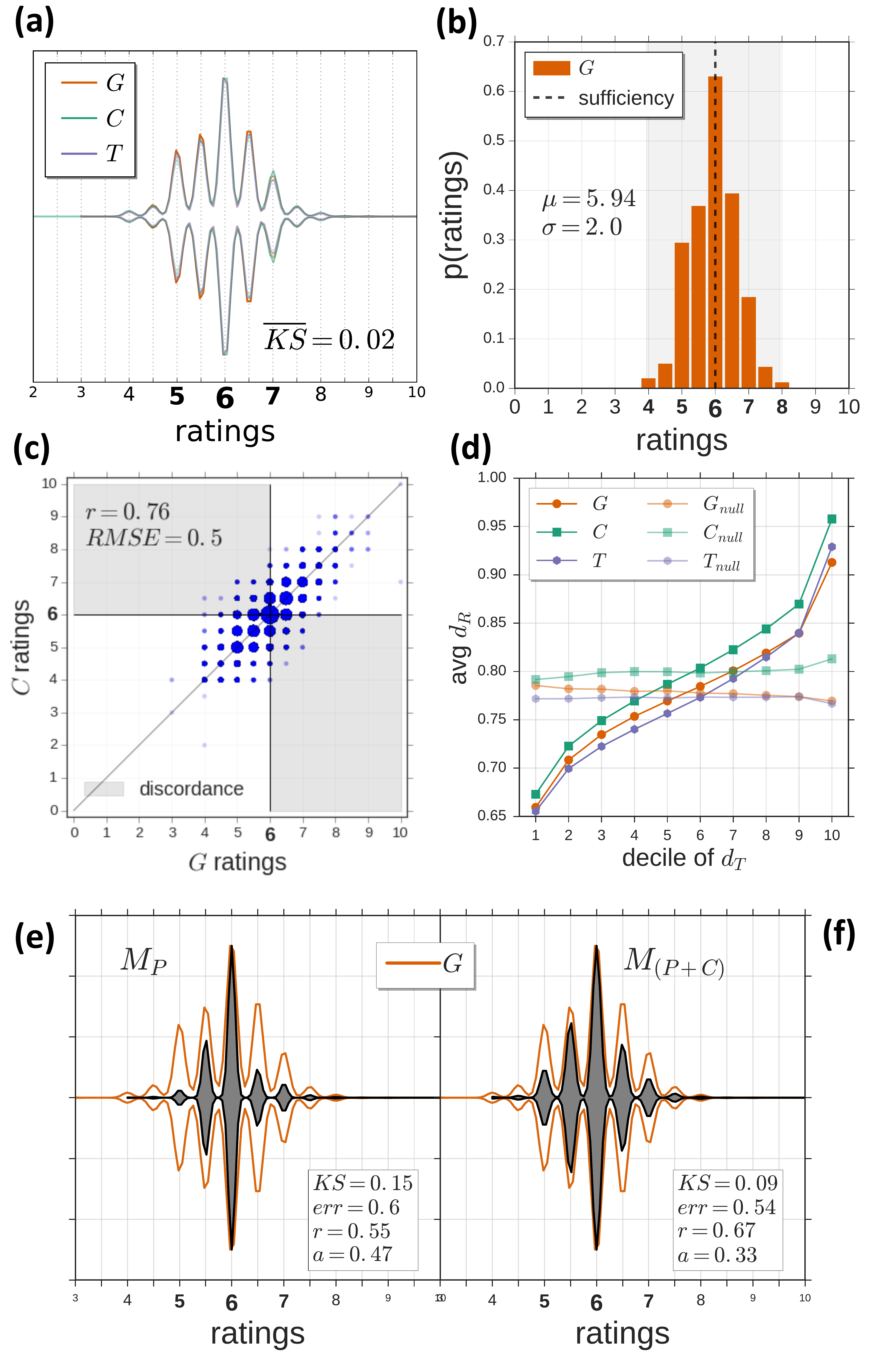}
    \caption{\tiny \textbf{The patterns of soccer player ratings.} \textbf{(a)} Violin plots comparing the distribution of ratings for the three newspapers. We see that they are almost identical, with an average Kolmogorov-Smirnoff statistics $\overline{KS} = 0.02$. \textbf{(b)} Distribution of ratings for newspaper $G$. The ratings are peaked around 6 which corresponds to sufficiency in the Italian high school rating system. Most of the ratings are between 4 and 8, while ratings below 4 and above 8 are extremely rare. \textbf{(c)} Correlation between newspapers $G$ and $C$. The same strong correlation ($r=0.76$) is observed for every pair of newspapers. \textbf{(d)} Decile of Minkowski distance $d_T$ between performances versus the the average absolute difference of ratings $d_R$ for the three newspapers ($G$, $C$ and $T$) and three null models where the ratings are randomly shuffled across the performances ($G_{null}$, $C_{null}$, $T_{null}$). \textbf{(e)} Comparison between the distributions of $M_P$ and $G$, and between the distribution of $M_{(P + C)}$ and $G$ \textbf{(f)}. We see that $M_{(P+C)}$ is closer to $G$ than $M_P$ is.}
    \label{fig:ratings}
\end{figure}

Given a set $P = \{\vec{p}_1, \dots, \vec{p}_n\}$ of technical performances and the corresponding list of a judge $R$'s ratings $\vec{v}_R = \{v_1, \dots, v_n\}$, with $R \in \{G, C, T\}$, we compute function $f$ by training a machine learning classifier $M_P$ which learns the relations between technical performance and a human judge's ratings (see Supplementary Note 5). We evaluate $M_P$'s ability of simulating $\mathcal{F}$ as its ability to produce a rating set which agrees with a human judge as much as two human judges agree with each other. 
Figure \ref{fig:ratings}e shows the statistical agreement between the ratings $\vec{v}_{M_P}$ assigned by $M_P$ and the ratings $\vec{v}_G$ assigned by newspaper $G$. We find that the root mean squared error indicating the quality of $M_P$'s prediction is $RMSE_P = 0.60$, while the Pearson correlation between $M_P$'s predictions and real ratings is $r_P = 0.55$ (Figure \ref{fig:ratings}e). In contrast, the statistical agreement between two human judges is characterized by $RMSE_P = 0.50$ and $r_P = 0.76$ (see Figure \ref{fig:ratings}c-d). 
Moreover, we find that the distance between the distribution of $\vec{v}_{M_P}$ and the distribution of $\vec{v}_R$, as expressed by the Kolmogorov-Smirnov statistics \cite{justel1997multivariate}, is $KS_P = 0.15$, while for two human judges $KS_P = 0.02$. On one hand, these results indicate that a human judge partly relies on technical features, since the function $f$ inferred by the training task is able to assign ratings with some degree of agreement with human judges. Indeed, a null model where ratings are assigned randomly according to the distribution of real ratings results in null correlations (see Supplementary Note 6). On the other hand, the disagreement indicates that the technical features alone cannot fully explain the rating process. This can be due to the fact that soccer logs do not capture some important aspects of technical performance, or by the fact that the judges consider aspects other than a player's technical performance, i.e., personal bias or external contextual information. 

To investigate the second hypothesis we retrieve contextual information about players, teams and games, such as age, nationality, the club of the player, the expected game outcome as estimated by bookmakers, the actual game outcome and whether a game is played home or away (see Supplementary Note 7). We then generate for every player an extended performance vector which include both technical and contextual features and train a machine learning classifier $M_{(P + C)}$ on the set of extended performances. Figure \ref{fig:ratings}f shows that, by adding contextual information, the statistical agreement between the artificial judge and the human judge increases significantly. In particular, the artificial judge $M_{(P + C)}$ is much more in agreement with human judge $G$ than $M_P$ is, with the correlation increasing from $r_{P}=0.56$ to $r_{(P + C)}=0.68$ and the root mean squared error decreasing from $RMSE_P = 0.60$ to $RMSE_{(P + C)} = 0.54$ (Figure \ref{fig:ratings}f). Furthermore, the distribution of $\vec{v}_{M_{(P + C)}}$ is more similar to $\vec{v}_G$ than $\vec{v}_{M_P}$ is, since $KS_{P} = 0.15$ while $KS_{(P + C)} = 0.09$ (Figure \ref{fig:ratings}f). 
We observe that the highest errors of $M_{(P + C)}$ are associated with the outliers (ratings $> 7$ or $< 5$) and can be due to either the scarcity of examples in the dataset or by the absence of other contextual knowledge. For example, 10 is assigned only two times in the dataset. One case is when Gonzalo Higua\'in, by scoring three goals in a game, became the best seasonal top scorer ever in the Italian Serie A. This is a missing variable which is not related to the player's performance in that game, making this exceptional rating impossible to predict. Artificial judge $M_{(P + C)}$ classifies that performance simply as a good one assigning to the player a high rating (8).  
Our results clearly show that $\mathcal{F}$ cannot be accurately inferred from technical performance only, as human observers rely also on contextual factors which go beyond the technical aspects a player's performance. 
For example, a victory by a player's team or an unexpected game outcome compared to the expectation recorded by bookmakers, significantly changes a player's rating, regardless of his actual technical performance. 

\subsection{Factors influencing human perception of performance (Explain).} 
\begin{figure}\centering
\includegraphics[scale=0.45]{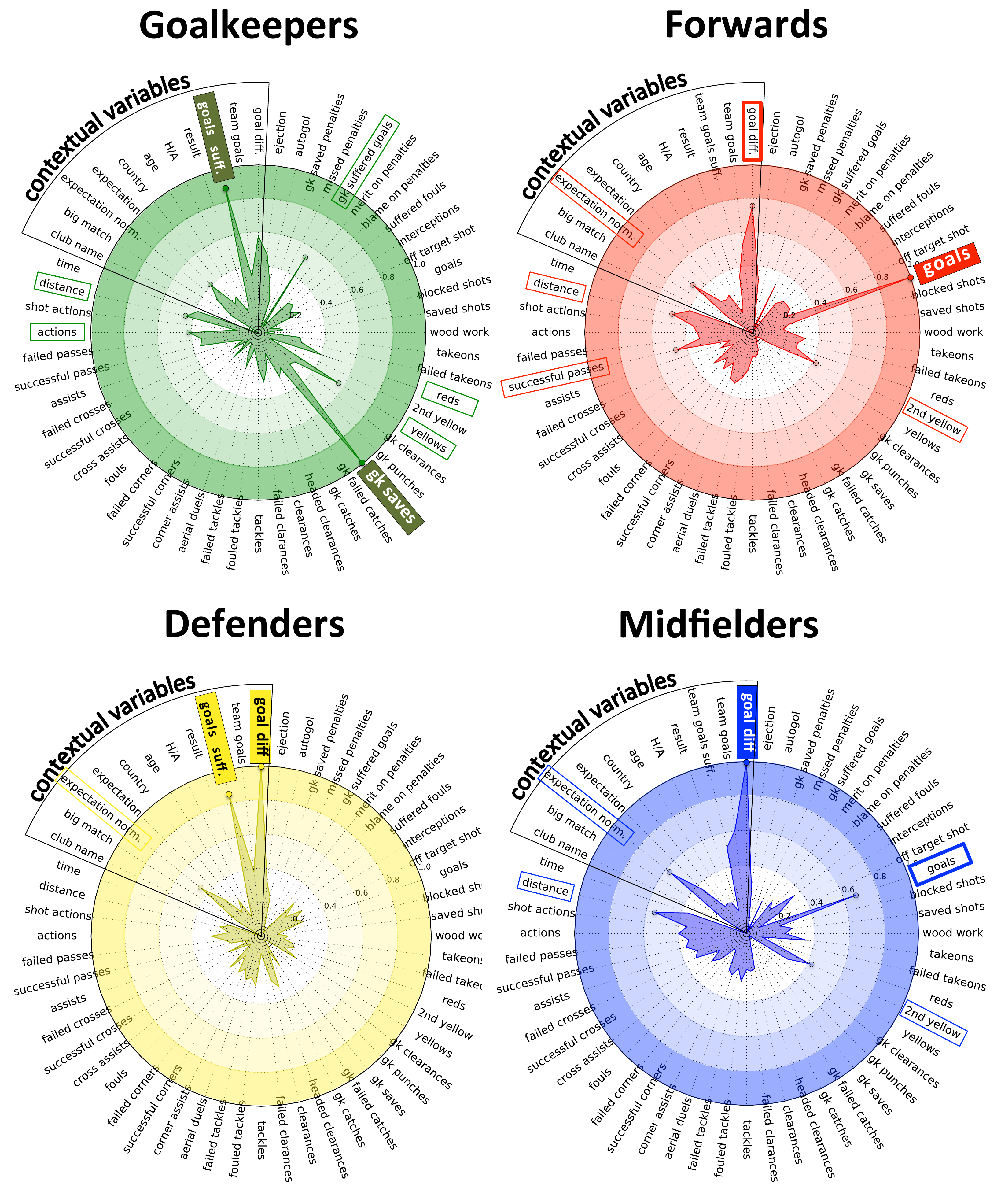}
\caption{\tiny \textbf{Importance of technical and contextual features to human rating process.} The radar charts indicate the importance of every feature, normalized in the range $[0, 1]$, to human rating process for Goalkeepers (a), Defenders (c), Midfielders (d) and Forwards (b). Feature with an importance $\ge 0.4$ are highlighted and contextual features are grouped in the left upper corner of every radar chart. We compute a feature's importance by a combinations of the weights produced by a repertoire of machine learning models (see Supplementary Note 9). The plots indicate that, for example, for forwards three features matter the most: the number of \emph{goals} scored (performance), the game goal difference (contextual) and the expected game outcome as estimated by bookmakers before the match (contextual).}
\label{fig:importances}
\end{figure}

Figure \ref{fig:importances} summarizes the importance of each technical and contextual feature to a human judge's rating process (Supplementary Note 8). 
As in soccer a player is assigned to one of four roles -- goalkeeper, defender, midfielder, forward -- each implying specific tasks he is expected to achieve, the importance of the features to a judge varies from role to role. For this reason, we compute the importance of technical and contextual variables for each role separately (Supplementary Note 8).

We observe that most of a human judge's attention is devoted to a small number of features, and the vast majority of technical features are poorly considered or discarded during the evaluation process. We further investigate this aspect by selecting the $k$ most important features for every $k=1, \dots, 150$ and evaluating the accuracy of $M_{(P + C)}$ on that subset of selected features. We find that the predictive power of $M_{(P + C)}$ stabilizes around 20 features, confirming that the majority of technical features have negligible influence to the human rating process (Supplementary Note 9).

We find that human perception of a feature's importance changes with the role of the player, i.e., the same feature has different importance for different roles. For example, while the most attractive feature for goalkeepers and forwards is a technical feature (saves and goals, respectively), midfielders and defenders attract a human judge's attention mainly by collective features like the team's goal difference. 
This is presumably because goalkeepers and forwards are directly associated with events which naturally attract the attention of observers such as goals, either scored or suffered. In contrast, the evaluation of defenders and midfielders is more difficult, as they are mainly involved in events like tackles, intercepts and clearances, which attract less the attention of human observers. Indeed, as Figure \ref{fig:importances} shows, for defenders and midfielders the contextual features have the highest importance, in contrast with goalkeepers and forwards which are mainly characterized by technical features (Figure \ref{fig:importances}).

\subsection{Noticeability heuristic.}
\begin{figure}\centering
\includegraphics[scale=0.34]{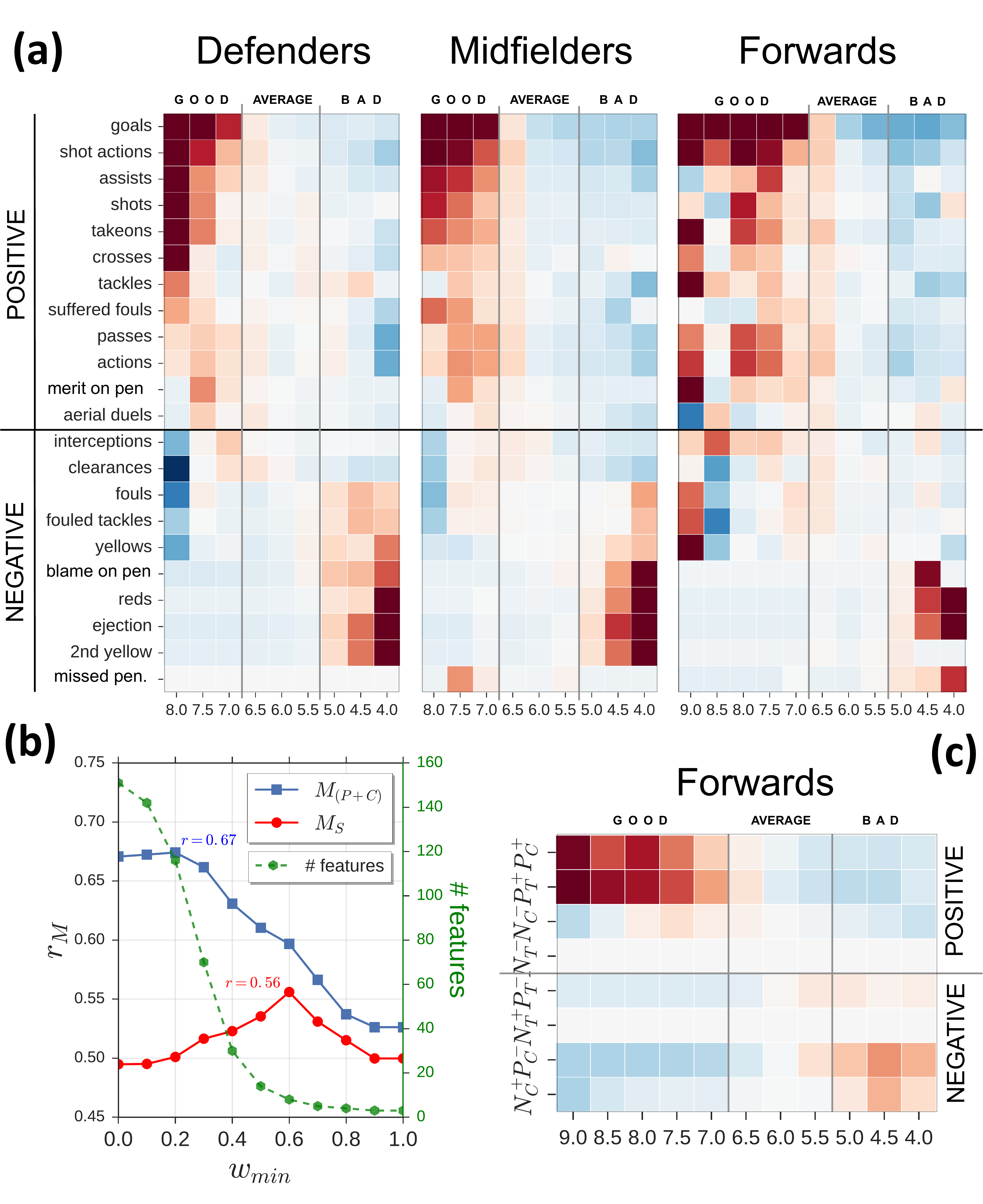}
\caption{\tiny (a) Typical performances for every rating, grouped by role on the field. The color of each cell, in a gradient from blue to red, represents the difference from the mean of the corresponding feature. Red cells indicate values above the average, blue cells indicate values below the average, white cells indicate values on the average of the feature's distribution. We observe that good ratings (> 6.5) are characterized by values of positive features above the mean and values of negative features below the mean. The opposite applies for bad ratings (< 5.5). (b) Correlations $r_{M_{(P +C)}}$ (blue curve) and $r_{M_S}$ (red curve) as less features are selected during the learning phase by imposing a minimum feature importance. The black curve indicates the number of features selected by a given minimum feature importance. Model $M_S$ has a Pearson correlation $r=0.55$ when aggregating over the 8 most important features, achieving the same predictive power as model $M_P$ which uses 150 technical features. (c) Relative importance of features in the noticeability heuristic model, for forwards (red), midfielders (blue), defenders (yellow) and goalkeepers (green). In the noticeability model, the evaluation of forwards and midfielders is dominated by positive technical and contextual features; for goalkeepers and defenders it is instead dominated by the number of negative contextual features.}
\label{fig:heatmap}
\end{figure}

Here we investigate how human judges use the available features to construct a performance's evaluation. Given a rating $r$, we define its average performance as $\vec{p}(r) = [\overline{x}^{(r)}_1, \overline{x}^{(r)}_2, \dots, \overline{x}^{(r)}_n]$, where $\overline{x}^{(r)}_i {=} \frac{1}{N} \sum_{u, g} x_i(u, g)$ is the average value of technical feature $x_i$ computed across all performance vectors associated with $r$. 
Figure \ref{fig:heatmap}a visualizes $\vec{p}(r)$ discriminating between positive features, which have a positive correlation with soccer ratings, and negative features which have a negative correlation with them. We observe that, for $r \in \{5.5, 6.0, 6.5\}$, feature values in $\vec{p}(r)$ are all close to the average, while for the other ratings features are either significantly higher (red) or lower (blue) than the average (Figure \ref{fig:heatmap}a). In other words, while performances associated with extreme ratings have some features with extreme values hence capturing a judge's attention, performances associated with ratings in the vicinity of 6 are characterized by average feature values, hence representing performances less worthy of attention. We quantify this aspect by defining a rating's \emph{feature variability} $\sigma(r)$ as the standard deviation of $\vec{p}(r)$. We find that extreme ratings are characterized by the highest values of $\sigma(r)$ (see Supplementary Note 10), denoting the presence in $\vec{p}(r)$ of some feature values far from the average. 
This suggests that a judge's evaluation relies on a \emph{noticeability heuristic}, i.e., they assign a rating to a performance based on the presence of feature values that can be easily brought to mind.

Driven by the above results, we label a feature value as \emph{noticeable} if it is significantly higher or lower than the average (see Supplementary Note 11).
We then aggregate the original feature space into a low-dimensional performance vector $S = [P_{T}^+, P_{T}^-, N_{T}^+, N_{T}^-, P_{C}^+, P_{C}^-, N_{C}^+, N_{C}^-]$, where $P$ indicates the number of positive features having noticeable values, $N$ indicates the number of negative features having noticeable values, symbols $+$ and $-$ indicate that the features considered are positive or negative respectively, $T$ and $C$ refer to technical and contextual features respectively. 
For each performance, we construct $S$ and train a machine learning model $M_S$ to predict soccer ratings from it. Figure \ref{fig:heatmap}b shows how $r_{M_{(P + C)}}$ (blue curve) and $r_{M_S}$ (red curve) changes as we reduce the number of features used during the learning phase by imposing a minimum feature importance $w_{min}$, which is a proxy of a feature's relevance to a judge's evaluation.
We find that while $r_{M_{(P + C)}}$ decreases as $w_{min}$ increases, $r_{M_S}$ has an initial increasing trend and reaches $r_{M_S} = 0.56$ when $w_{min} = 0.6$, equalizing artificial judge $M_P$. Notably, when $w_{min} = 0.6$ only 8 features out of 150 are selected to construct $S$, corresponding to the features in the two widest circles in the radar charts of Figure \ref{fig:importances}. The presence of a peak at $w_{min} = 0.6$ in Figure \ref{fig:heatmap}b suggests how a human judge constructs the evaluation: first she selects the most relevant features and then verifies the presence of noticeable values in them.

Figure \ref{fig:heatmap}c visualizes the average performance $\vec{p}_*(r) = [\overline{P}_{T}^+, \overline{P}_{T}^-, \overline{N}_{T}^+, \overline{N}_{T}^-, \overline{P}_{C}^+, \overline{P}_{C}^-, \overline{N}_{C}^+, \overline{N}_{C}^-]$ consisting of the average values computed across all the low-dimensional performance vectors associated with a rating $r$. We find that average performances associated with extreme ratings are characterized by the presence of noticeable feature values (Figure \ref{fig:heatmap}c). In particular, for forwards and midfielders the attention of the judges is dominated by positive technical and contextual features $P_{T}^+$ and $P_{C}^+$, which include the most prominent features like goals and the game's goal difference (Supplementary Figure 6). Defenders are instead dominated by the number of negative contextual features $N_{C}^-$, which include for example the number of goals suffered (see Supplementary Figure 6). Overall, the description of performance based on the noticeability heuristic has the same predictive power as a description using 150 technical features. These results clearly show the simplicity of the human rating process, indicating that (i) human judges select only a few, mainly contextual, features and (ii) they base the evaluation on the presence of noticeable values in the selected features. 

\section*{Conclusion}

In this work we investigated the cognitive process underlying performance evaluation. We find that human judges select a small set of indicators which capture their attention. Contextual information is particularly important to the human evaluation process, indicating that similar technical performances may have different evaluations according to a game's results or outcome expectation. 
Moreover, we discover that human judges rely on a noticeability heuristic, i.e., they select a limited number of features which attract their attention and then construct the evaluation by counting the number of noticeable values in the selected features.
The analytical process proposed in our study is a first step towards a framework to ``evaluate the evaluator'', a fundamental step to understand how human evaluation process can be improved with the help of data science and artificial intelligence. Although in many environments evaluations as objective as possible are desirable, it is essential that human evaluators inject their values and experience into their decisions, possibly taking into account aspects that may not be systematically mirrored in the available data. To this extent, the observe-predict-explain process exemplified in this paper can be used to empower human evaluators to gain understanding on the underlying logic of their decisions, to assess themselves also in comparison with peers, and to ultimately improve the quality of their evaluations, balancing objective performance measurements with human values and expertise.

\newpage
\bibliographystyle{naturemag}
\bibliography{biblio}

\begin{addendum}
 \item We thank Daniele Fadda for his invaluable support on data visualization, Alessio Rossi and Marco Malvaldi for their useful suggestions. This work has been partially funded by the EU project SoBigData grant n. 654024.
 \item[Competing Interests] The authors declare that they have no
competing financial interests.
 \item[Correspondence] Correspondence and requests for materials
should be addressed to Luca Pappalardo~(email: lpappalardo@di.unipi.it,  lucapappalardo1984@gmail.com) and Paolo Cintia~(email: paolo.cintia@gmail.com).
\end{addendum}

\renewcommand{\figurename}{Supplementary Figure}
\renewcommand{\tablename}{Supplementary Table}
\renewcommand{\refname}{Supplementary References} 

\newpage
\section*{Supplementary Notes}
\subsection{Supplementary Note 1: Soccer Logs.}
Table \ref{tab:data} shows a sequence of events that occurred during the match AS Roma vs Juventus FC. Since the size of the football pitch is slightly variable from stadium to stadium -- UEFA establishes that it can be 100 to 105 meters long, and 64 to 68 meters wide\footnote{\url{https://en.wikipedia.org/wiki/UEFA_stadium_categories}} -- we normalize both the field coordinates in the range $[0, 100]$. The colored row in Table \ref{tab:data} shows an example of event, where player Totti (AS Roma) makes a successful pass from position $(55.3, 23.4)$ of the field 1,389 seconds into the game. 

\begin{table}\centering
\begin{tabular}{|c|c|c|c|c|c|}
\hline
\textbf{team} & \textbf{player} & \textbf{event} & \textbf{outcome} & \textbf{where} & \textbf{time}\\
\hline
\vdots &\vdots &\vdots &\vdots &\vdots &\vdots\\ 
Roma & De Rossi & pass & success & (68.9,55) &  1253\\
Roma & Totti & attempt & failed & (88.4,60.3) & 1343\\
\cellcolor{blue!20}Roma & \cellcolor{blue!20}Totti & \cellcolor{blue!20}pass & \cellcolor{blue!20}success & \cellcolor{blue!20}(78.3,40.2) & \cellcolor{blue!20}1389\\
Juventus & Pogba & pass & failed & (46.3,17.8) & 1406\\
Roma & Salah & tackle & success & (59.5,86.1)  & 1409\\
Juventus & Dybala & pass & success & (62.8,67.5) & 1416\\
\vdots & \vdots &\vdots &\vdots &\vdots &\vdots \\
\hline
\end{tabular}
\caption{\small \textbf{Example of soccer logs.} Example of events that occurred during the match AS Roma vs Juventus FC. Each event has the following fields: (i) the team who generated the event; (ii) the player who generated the event; (iii) the type of event; (iv) the outcome of the event; (v) the position on the field where the event originated; (vi) the time-stamp of the event (in seconds since the beginning of the match).}
\label{tab:data}
\end{table}

\subsection{Supplementary Note 2: Technical Features.}

The soccer log dataset consists of a collection of events that occurred during the matches in the Italian Serie A in seasons 2015/2016 and 2016/2017. Each event is characterized by spatial information (expressed in relative coordinates: $x \in [0,100], y \in [0,100] $ ), time information (expressed in seconds since the beginning of the game) and outcome information (whether or not the event is successful). 
In particular, the available event types are:
\begin{itemize}
\item \emph{pass}: a player passes the ball to a teammate. A pass is successful if it reaches the teammate, and failed otherwise. An assist is a special case of successful pass where the teammate receiving the ball consequently scores a goal.
\item \emph{tackle}: a player tries to take the ball from an opponent. A tackle is successful is the player successfully takes the ball, or failed otherwise. 
\item \emph{dribble} (or take-on): a player holding the ball avoids the tackle of an opponent. A dribble is successful if the player successfully avoids the tackles of the opponent, or failed otherwise.
\item \emph{headed duel}: a player faces an opponent on gaining the possession of the ball with a head shot, while the ball is floating up in the air.
\item \emph{cross}: a player makes a cross when he throws the ball inside the opponents' penalty area. In practice, it is a particular type of pass. 
As for passes, a cross can be successful, failed or assist. 
\item \emph{shot} (or goal attempt): a player performs a shot towards the opponents' goal. A shot can be either off target, saved (if the goalkeeper blocks it), blocked (if an opponent blocks the ball before it reaches the goalkeeper or the goal line), or a goal.
\item \emph{intercept}: a player intercepts a pass of an opponent and recovers the ball. 
\item \emph{clearance}: a player throws the ball far away from his penalty area. A clearance can be \emph{successful} or \emph{failed}. It is successful if the ball is recovered by teammates, it is failed if the ball is recovered by the opponents.
\item \emph{goal keeping}: the goalkeeper saves, intercepts or blocks a shot from an opponent. 
\item \emph{foul}: a player performs an action that is against the rules. A foul can be active, if the player commits the foul, or passive if the player suffers a foul from an opponent.

\end{itemize}

Given a game $g$, we describe the performance $\vec{p}(A, g)$ of a player $A$ by aggregating the recorded events he generates in $g$. In particular, we define four categories of technical features extracted from soccer logs:
\begin{itemize}
\item \textbf{Quantity features}: for every event type, we count the number of events of that type generated by the player during a game. This produces the following features: 
    \begin{itemize}
        \item \emph{pass} - total number of passes 
        \item \emph{tackle} - total number of tackles 
        \item \emph{dribble} - total number of dribblings 
        \item \emph{headed duel} - total number of headed duels
        \item \emph{cross} - total number of crosses 
        \item \emph{shot} - total number of shots 
        \item \emph{intercept} - total number of intercepts 
        \item \emph{clearance} - total number of clearances
        \item \emph{goal keeping} - total number of goals keeping saves
        \item \emph{foul} - total number of fouls 
        \item \emph{card} - total number of cards (yellows or reds)
        \item \emph{minutes played} - the total time (in minutes) the player has played during the game
        \item \emph{distance traveled} - the total distance the player has traveled during the game
    \end{itemize}
\item \textbf{Quality features}: for every event type, we count the number of successful events (e.g., successful pass, successful tackles, etc.) and the number of failed events (e.g., failed passes, failed tackles, etc.) of that type. This produces the following features: 
    \begin{itemize}
        \item \emph{completed pass} - total number of completed passes 
        \item \emph{failed pass} - total number of failed passes 
        \item \emph{key pass} - total number of key passes. A key pass is a pass leading to a shot of the player who received the ball
        \item \emph{assist pass} - total number of assists. An assist is a completed pass to a player who consequently scores a goal
        \item \emph{completed cross} - total number of completed crosses 
        \item \emph{failed cross} - total number of failed crosses 
        \item \emph{assist cross} - total number of cross assists. 
        \item \emph{completed tackle} - total number of completed tackles 
        \item \emph{failed tackle} - total number of failed tackles
        \item \emph{fouled tackle} - total number of fouled tackles. A tackle is fouled when its outcome is a foul
        \item \emph{completed dribble} - total number of completed dribblings
        \item \emph{failed dribble} - total number of failed dribblins.
        \item \emph{completed headed duel} - total number of completed headed duels
        \item \emph{failed headed duel} - total number of failed headed duels
        \item \emph{fouled headed duel} - total number of fouled completed headed duels. An headed duel is fouled when its outcome is a foul
        \item \emph{intercept} - total number of intercepts
        \item \emph{completed intercept} - total number of completed intercepts 
        \item \emph{completed clearance} - total number of completed clearances
        \item \emph{failed clearance} - total number of failed clearances
        \item \emph{headed clearance} - total number of headed clearances
        \item \emph{off-target shot} - total number of completed dribblings 
        \item \emph{saved shot} - total number of shots saved by the opponent goal keeper
        \item \emph{blocked shot} - total number of blocked shots. A shot is blocked when an opponent intercepts the ball before it reaches the goal keeper
        \item \emph{completed shot (goal)} - total number of goals scored 
        \item \emph{wood work shot} - total number of shots hitting the post
        \item \emph{own goal} - total number of own goals
        \item \emph{missed penalty} - total number of missed penalties
        \item \emph{scored penalty} - total number of scored penalties
        \item \emph{blame on penalty} - total number of fouls whose outcome is a penalty for the opponents
        \item \emph{merit on penalty} - total number of suffered fouls whose outcome is a penalty for player's team
        \item \emph{save goal keeping} - total number of saves as goalkeeper
        \item \emph{catch goal keeping} - total number of catches as goalkeeper
        \item \emph{punch goal keeping} - total number of punches as goalkeeper
        \item \emph{failed catch goal keeping} - total number of failed catches as goalkeeper
        \item \emph{clearance goal keeping} - total number of clearances as goalkeeper
        \item \emph{suffered goal goal keeping} - total number of goals suffered as goalkeeper
        \item \emph{penalty saved} - total number of penalty saved as goal keeper
        \item \emph{foul made} - total number of fouls made
        \item \emph{foul suffered} - total number of fouls suffered
        \item \emph{yellow card} - total of yellow cards
        \item \emph{red card} -  total number of red cards
        \item \emph{2nd yellow card} - total number of second yellow cards
        \item \emph{ejection} - total number of ejections (second yellow cards or red cards)
        \item \emph{total action participation}: the total number of actions the player has been involved in. An action is a sequence of events where the same team is in control of the ball. 
        \item \emph{successful action participation}: the total number of successful actions the player has been involved in. A successful action is an action ending with a goal attempt (shot).
    \end{itemize}
    
\item \textbf{Contribution features}: given the value $q(A, g)$ of a quality feature $q$ for a player $A$ of team $T$ in game $g$, the value of the corresponding contribution feature is: 
$$c_q(A, g) = \frac{q(A, g)}{\sum_{i \in E_T} q(i, g)},$$
where $q(i, g)$ is the value of quality feature $q$ for player $i$ in game $g$ and $E_T$ is the set of players of team $T$. A contribution feature quantifies a player's contribution to the performance of his team with respect to a quality feature. For example, if a player $A$ scores in game $g$ one out of four goals scored by his team, the value of the corresponding contribution feature is $c_{goals}(A, g) = 0.25$. If $A$ instead scores all the four goals, his contribution features will be $c_{goals}(A, g) = 1.0$ indicating that the player's contribution to the team is much stronger. 

\item \textbf{Dangerousness features}: we weight every quality feature according to the probability of scoring a goal from the location where the corresponding event is generated. To compute these features, we associate a weight to each event according to Figure \ref{fig:dangerousness}. Given the value $q(A, g)$ of a quality feature $q$ for a player $A$, the corresponding dangerousness feature is computed as: $$d_q(A,g)=\sum_{i \in q(A,g)} w(i_x,i_y), $$ where $w$ is the heatmap function in Figure \ref{fig:dangerousness}, and
$i_x,i_y$ are the coordinates on the pitch where event $i$ occurred. Note that there is a distinction between completed and failed events. For example, a completed pass close to opponents' goal has a greater value than the same pass performed far from the opponents' goal. On the other hand, a failed pass is more dangerous when it occurs close to the player's own goal. To model this fact, for events where opponent team gets the ball (e.g. failed pass, failed tackles, etc), the point $i_x,i_y$ is recomputed as $i'_x,i'_y=100-i_x,100-i_y$. Since the coordinates of events are expressed on a 0-100 scale, with this operation a failed event is weighted more if it occurs close to player's own goal, conversely to what happens with completed events. 

\end{itemize}

\begin{figure}\centering
\includegraphics[scale=0.5]{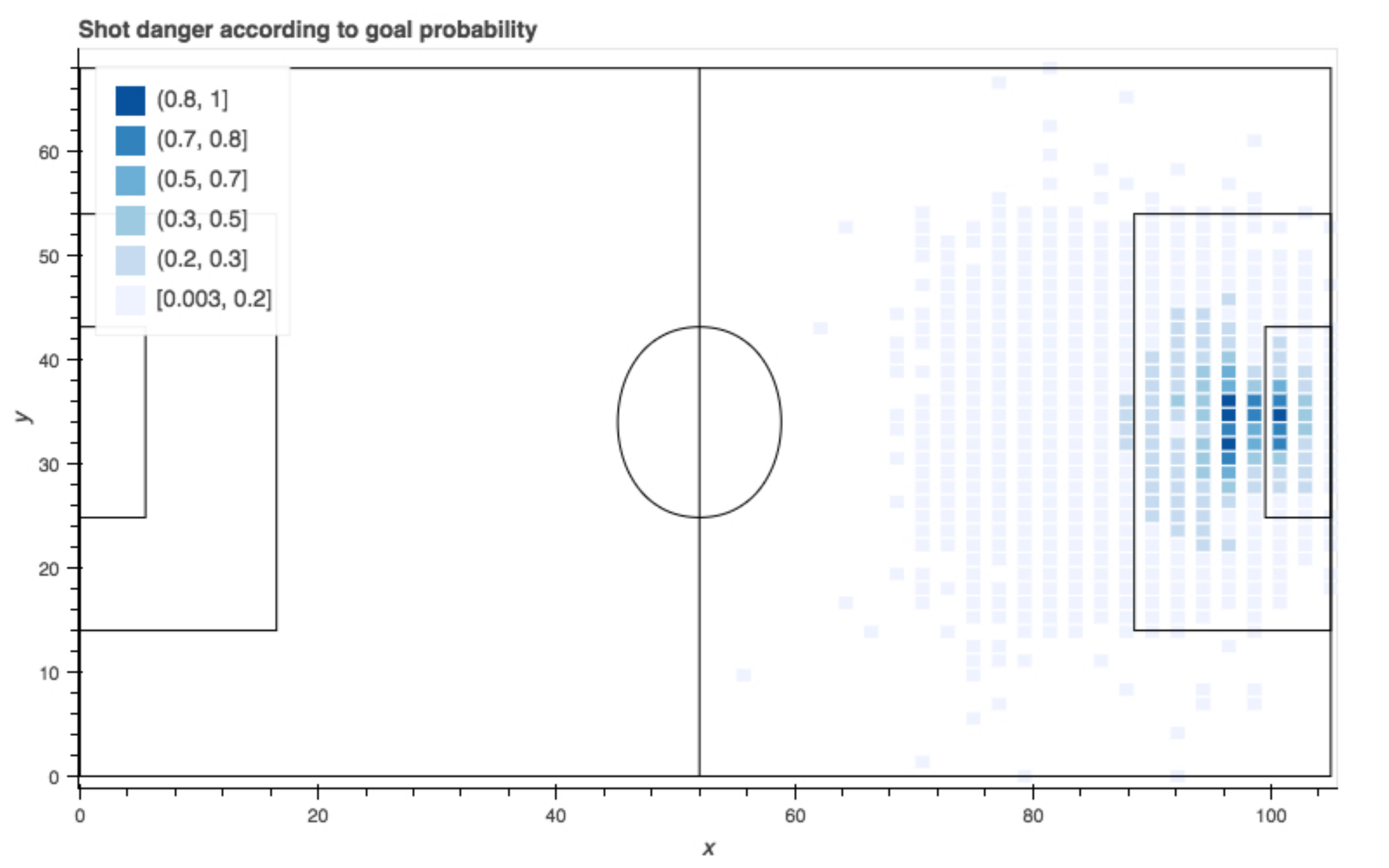}
\caption{Heatmap of the dangerousness of events in soccer games. The color of cells, in a gradient from light blue to dark blue, indicates the dangerousness of each zone of the field, computed by aggregating the positions of all shots that converted into goals.}
\label{fig:dangerousness}
\end{figure}

\subsection{Supplementary Note 3: Bland-Altman plot.}
Bland and Altman (BA) plots are used to evaluate the agreement among two different measurements of the same phenomenon \cite{altman1983methods, bland1986statistical, giavarina2015understanding}. In a BA plot the difference of two paired measurements is plotted against their mean. The mean difference between the measurements is the estimated bias: if this value differs significantly from 0, this indicates the presence of fixed bias. The standard deviation of the differences measures the random fluctuations around the mean: if the differences within mean $\pm 1.96$SD are not important, the two ratings may be used interchangeably. 

\begin{figure}
    \centering
    \includegraphics[scale=0.5]{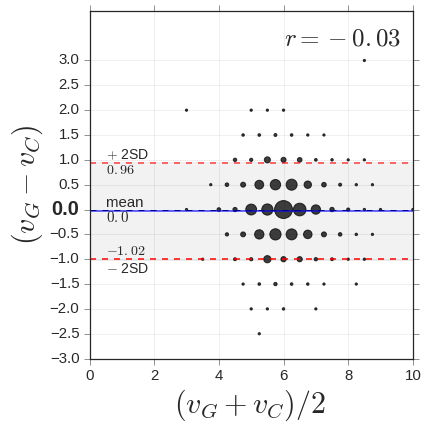}
    \caption{Scatter plot of mean versus difference of paired ratings for newspapers $G$ and $C$ (Bland-Altman plot). The mean difference is zero indicating that there is no systematic bias between the two newspapers. The limits of agreement are $[−1, +1]$ indicating that 95\% of the differences between the paired ratings are lower than 1.}
    \label{fig:bland_altman}
\end{figure}

We use BA plot to evaluate the agreement among two newspapers and identify the presence of systematic bias. Figure \ref{fig:bland_altman} plots the mean of each rating by $G$ and $C$ versus the two ratings' difference. We find that (i) the mean difference is zero, meaning that there is no fixed bias between the two newspapers; and (ii) the standard deviation is small, confirming that the ratings compiled by $G$ and $C$ are statistically indistinguishable. We find similar results when comparing $G$ with $T$ and $C$ with $T$.

\subsection{Supplementary Note 4: Minkowski distance.}
It has been observed that in high dimensional spaces (more than 3 dimensions) the ratio of the Euclidean distances of the nearest and farthest neighbors to a given target in high dimensional space in almost 1 for a wide variety of data distributions and distance functions \cite{aggarwal2001}. Although many algorithms use the Euclidean distance metric as natural extension of its traditional use in two- or three-dimensional spatial applications, recent research demonstrates that in high dimension space (such as the case of soccer performance) using fractional distance metrics, i.e., metrics for which the $L_k$ norm has $k \in (0, 1]$, is more effective at preserving the meaningfulness of distance measures \cite{aggarwal2001}. The $L_k$ norm, also called Minkowski distance, is defined as: $$ L_k(x, y) = \sum_1^d (||x^i - y^i||^k)^{1/k}, $$ where $x, y \in \mathbb{R}^d$. As suggested by \cite{aggarwal2001}, in our study we use a fractional distance by imposing $k=0.5$.

\subsection{Supplementary Note 5: Construction of the artificial judge.}
We treat the construction of the artificial judge as a problem of ordinal classification (also known as ordinal regression), i.e., the problem of rating objects with values ranging on an ordinal scale \cite{Gaudette2009}. Soccer player ratings have indeed discrete values ranging from 0 to 10 with intervals of 0.5, which are referred in data mining as ordinal classes. Ordinal classification lies in-between single-label classification, in which the labels belong to an unordered set, and metric regression, where labels belong to a continuous set, typically the set $\mathbb{R}$ of real numbers.\footnote{We implement ordinal classification using the publicly available Python package \texttt{mord}, \url{http://pythonhosted.org/mord/}.} 

To build the artificial judge, we split the technical performances in four groups according to the role of the corresponding players (goalkeepers, defenders, midfielders, forwards), and construct a training dataset for every role. This choice is motivated by the fact that players with different roles are supposed to behave differently during a game, hence influencing human perception of performance in different ways. Given a role $r$, a training dataset $D_r$ is a set of examples (or instances) consisting of a technical performance vector $\vec{p}$ and a reporter's rating $v$. We train a different ordinal classifier $C_r$ for every dataset, and validate each classifier by using a cross-validation strategy with $k=10$. In 10-folds cross-validation, a dataset is split into ten parts: in turn, nine parts are used for training the model and the remaining part as a test set for validation. 

Every time a technical performance for a player with role $r$ has to be associated with a rating, the classifier $C_r$ is selected and used to obtain the rating.

\subsection{Supplementary Note 6: Null Model.}
\begin{figure}
    \centering
    \includegraphics[scale=0.5]{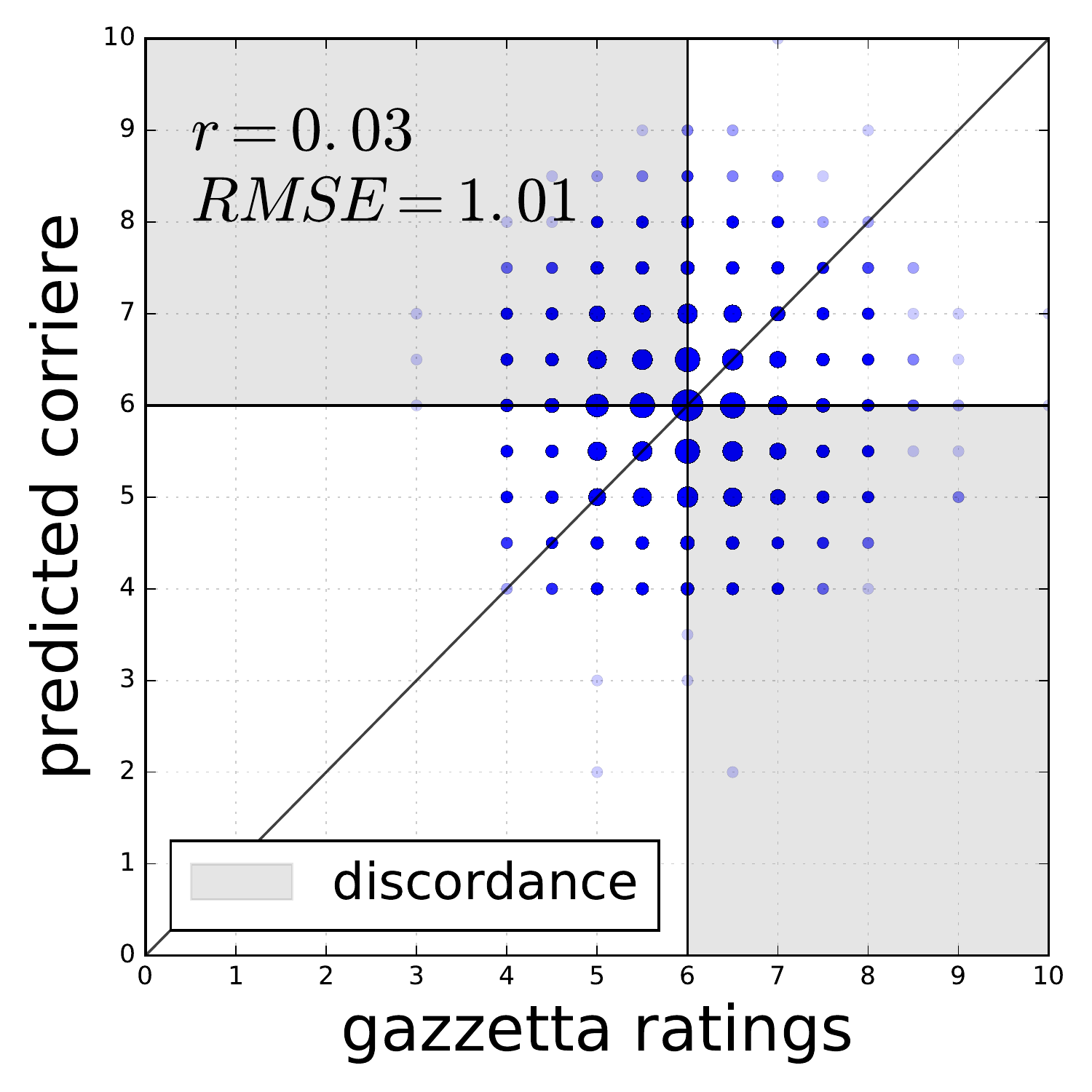}
    \caption{Correlation between ratings assigned by the null model $M_{null}$ and ratings assigned by judge $C$. We find that $r_{M_{null}} \approx 0.0$ and $RMSE_{M_{null}} \approx 1$, indicating that there is not statistical agreement between the artificial ratings assigned by $M_{null}$ and the ratings assigned by $C$.}
    \label{fig:null_model}
\end{figure}

We create a null model $M_{null}$ where ratings are assigned to performances randomly according to the distribution of ratings. We find that $r_{M_{null}} = 0.0$, i.e., there is no statistical agreement between the distribution of real ratings and the ratings assigned by $M_{null}$ (see Figure \ref{fig:null_model}). Moreover, $RMSE_{M_{null}} = 0.98$ indicating that, in average, the error between the artificial ratings and the human ratings are is about 2 ratings units, almost twice the error observed between two newspapers.

\subsection{Supplementary Note 7: Contextual features.}

Contextual features describe information about a player, a game or a team and are independent of a specific player's performance:

\begin{itemize}
\item \emph{Goal difference}: difference between the goals scored by the player's team and the goals scored by the opponents
\item \emph{Goals scored}: goals scored by the player's team
\item \emph{Goals suffered}: goals suffered by the player's team
\item \emph{Game outcome}: final outcome of the game for the player's team. Outcomes could be either \textit{Victory}, \textit{Defeat} or \textit{Draw}
\item \emph{Home/Away}: location of the game for the player's team. Could be either \textit{Home} for home games or \textit{Away} for away games.
\item \emph{Age}: Player's age
\item \emph{Country}: Player's nationality
\item \emph{Quote}: difference between the \textit{Ante-Post} odds of player's team and opponents. The higher such a difference the more a player's team is not expected to win the match. Ante Post football betting mainly concerns predicting the winner of the tournament before the tournament has started.
\item \emph{Big Match}: could be either \textit{True} or \textit{False} whether the game involves two team among the top 5 of the rankings or not.  
\end{itemize}

\subsection{Supplementary Note 8: Feature importance.}
To evaluate the importance of features to soccer player ratings we implement a repertoire of feature selection methods: Recursive Feature Selection (RFE), Stability Selection (SS) and Mean Decrease Impurity (MDI). A feature selection method evaluates the importance of the features as their predictive power in a classification, regression or ordinal classification model with respect to a target variable. We use the average of the importances estimate by the above three methods as an estimate of the feature's importance.

\paragraph{RFE (Recursive Feature Elimination).} \cite{guyon2002gene}. Recursive feature elimination (RFE) is based on the idea to repeatedly construct a predictive model and choose either the best or worst performing feature, setting the feature aside and then repeating the process with the rest of the features.
We repeatedly construct an ordinal classification model and choose the best performing feature based on the model's coefficients, setting the feature aside and then repeat the process with the rest of the features. The process is applied until all features in the dataset are exhausted. Features are then ranked according to when they were eliminated.\footnote{We use the RFE module in the Python library \texttt{scikit-learn}.}

\paragraph{SS (Stability Selection).} Stability selection (SS) is a method for feature selection based on subsampling in combination with feature selection algorithms \cite{meinshausen2010stability}.
It consists in applying a feature selection algorithm on different subsets of the data and with different subsets of the features. After repeating the process a number of times, we aggregate the selection results by checking how many times a feature ended up being selected as important when it was in an inspected feature subset. We expect important features to have scores close to 100\%, since they are always selected when possible. Weaker but still relevant features will also have non-zero scores, since they would be selected when stronger features are not present in the currently selected subset, while irrelevant features would have scores close to zero, since they would never be among selected features. We estimate the importance of features according to the SS paradigm by raking the SS scores computed by the Randomized Lasso method implemented in the Python package \texttt{scikit-learn}.

\paragraph{MDI (Mean Decrease Impurity).} When using a decision tree regressor, every node in the tree is a condition on a single feature, designed to split the dataset into two so that similar response values end up in the same set. The measure based on which the locally optimal condition is chosen is called
impurity, which in regression trees is typically its variance \cite{breiman1984classification}. Thus when training a tree, it can be computed how much each feature decreases the weighted impurity in a tree. We use the Regression Tree module provided by the Python package \texttt{scikit-learn} to implement the MDI feature selection method.

\subsection{Supplementary Note 9: Performance stability.}
\begin{figure}
    \centering
    \includegraphics[scale=0.5]{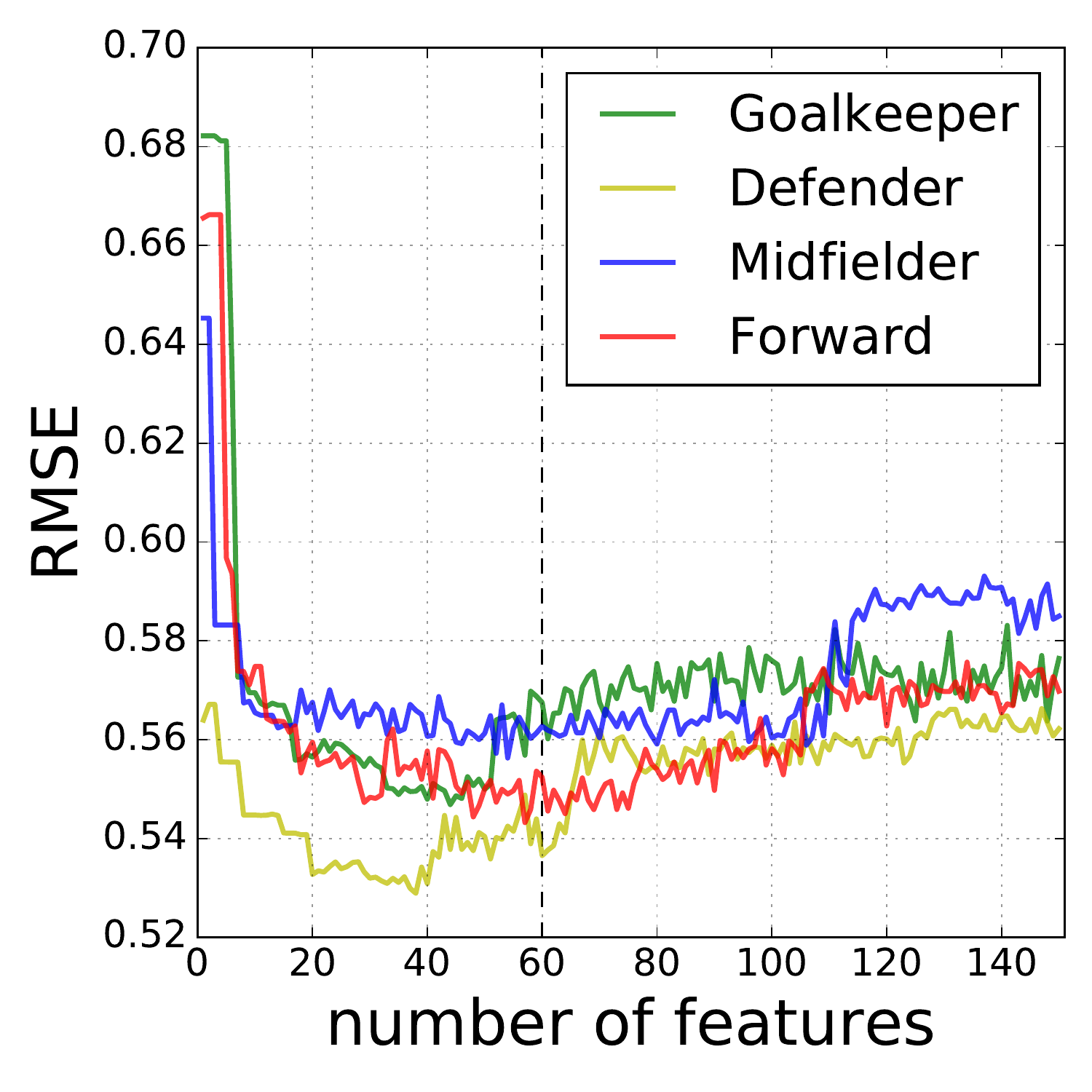}
    \caption{$RMSE$ of model $M_{(P + C)}$ as more features are considered to describe technical performance. 
    We observe that $RMSE_{M_{(P + C)}}$ stabilizes after 60 features.}
    \label{fig:performance_stability}
\end{figure}

We deeply investigate the impact of technical and contextual features to the human evaluation process by performing the following experiment. For each $k = 1, \dots, n$, where $n=150$ is the number of technical and contextual features considered in our study, we first select the $k$ features having the best ANOVA F-value w.r.t. the rating label and then train a model $M^*_{(P + C)}$ using only those features. Figure \ref{fig:performance_stability} shows how $RMSE_{M^*_{(P + C)}}$ changes as more and more features are considered. We find that $RMSE_{M^*_{(P + C)}}$ stabilizes around 20 features, indicating that most of the features have a negligible influence on the human judge's evaluation process.

\subsection{Supplementary Note 10: Feature variability.}
\begin{figure}
    \centering
    \includegraphics[scale=0.5]{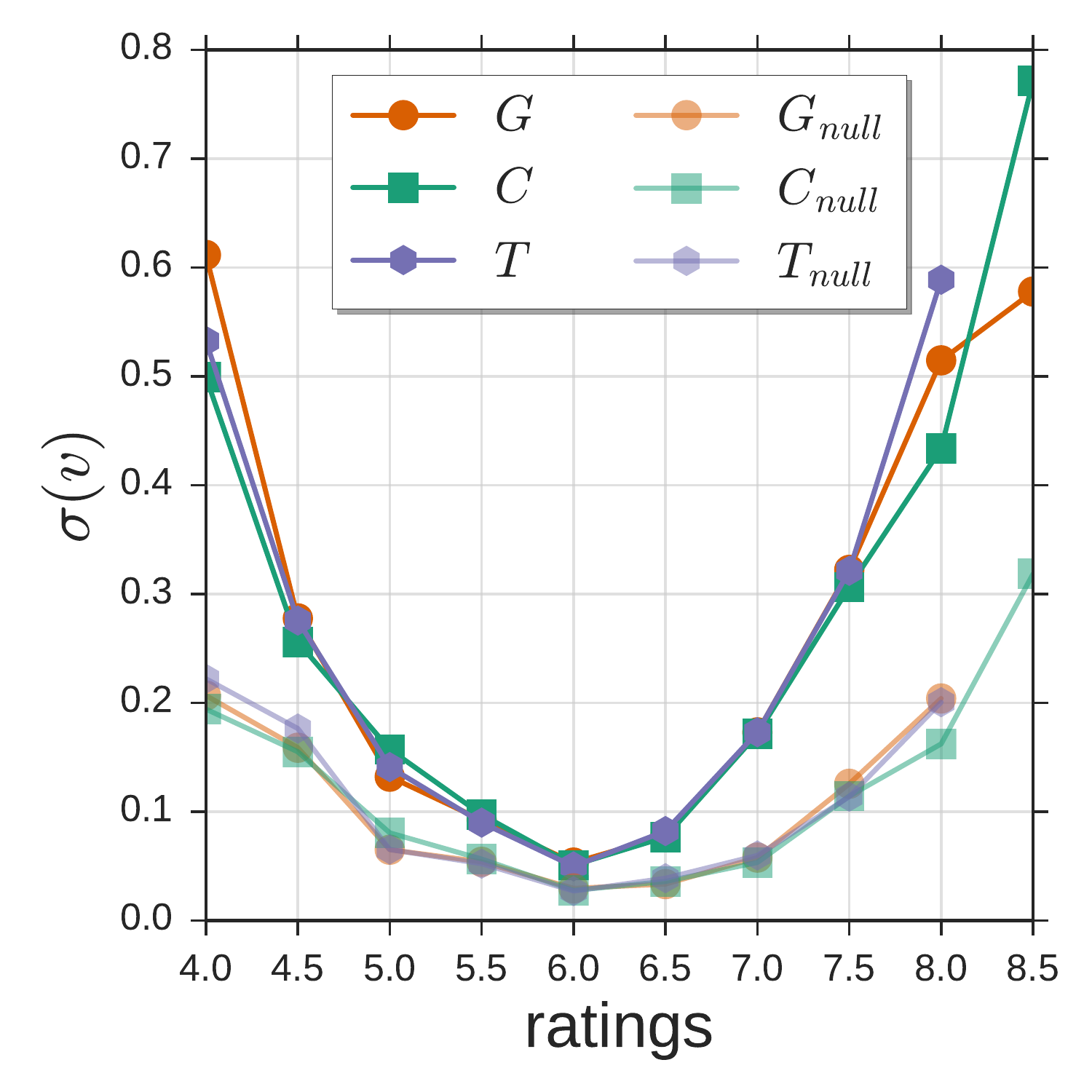}
    \caption{Soccer player ratings vs inter-feature variation. The plot shows the value of inter-feature variation $\sigma(v)$ for every rating $v$ assigned by judge $G$. We find that extreme ratings are characterized by the highest value of $\sigma(v)$.}
    \label{fig:inter_feature}
\end{figure}

We define a rating's feature variability $\sigma(r)$ as the standard deviation of the average performance $\vec{p}(r)$ of rating $r$. The higher $\sigma(r)$, the higher is the variance across the values in $\vec{p}(r)$. Figure \ref{fig:inter_feature} shows how the feature variability $\sigma(r)$ varies with the ratings. We find that the most frequent rating ($r=6$) is associated with the lowest feature variability $\sigma(6) \approx 0.05$, while extreme ratings are associated with the highest feature variabilities $\sigma(4) \approx 0.61$ and $\sigma(8.5) = 0.58$. We observe that null model $M_{null}$ (Supplementary Note 6) produces similar feature variability for rating close to the average, but fails in detecting the peaks associated with extreme ratings (Figure \ref{fig:inter_feature}).

\begin{figure}
    \centering
    \includegraphics[scale=0.4]{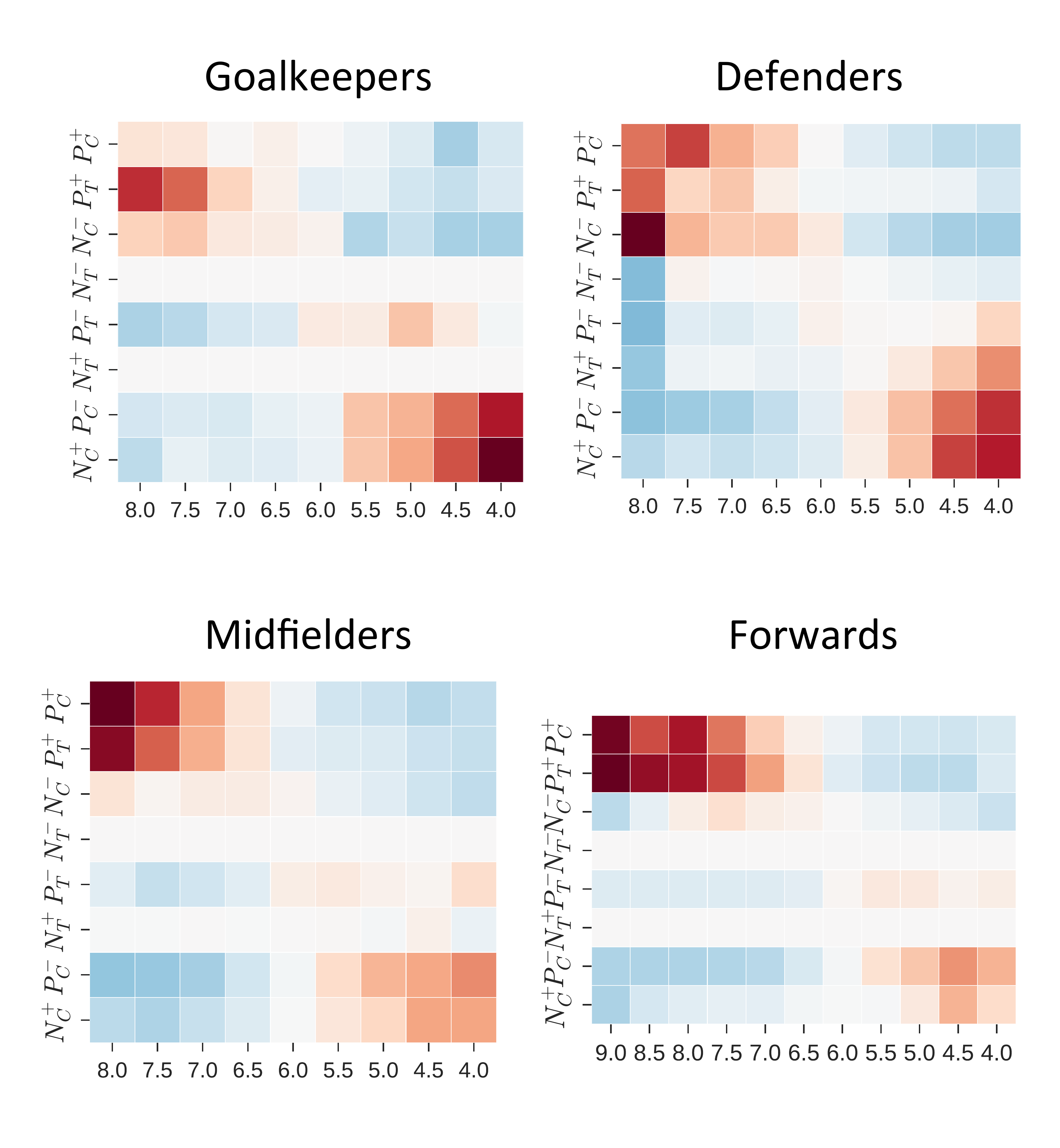}
    \caption{Noticeability heuristic: typical performance for every rating, grouped by role on the field. The color of each cell, in a gradient from blue to red, represents the difference from the mean of the corresponding feature. Red cells indicate values above the average, blue cells indicate values below the average, white cells indicate values on the average of the feature's distribution.}
    \label{fig:inter_feature2}
\end{figure}

\subsection{Supplementary Note 11: Noticeable features.}
A feature $x$ is positive if it has a positive correlation with soccer player ratings, otherwise it is a negative feature. Given the mean $\mu$ and the standard deviation $\sigma$ of the distribution of $x$, a feature value $z$ is significantly higher than the average if $z \ge \mu + 2\sigma$. Similarly, $z$ is significantly lower than the average if $z \le \mu - 2\sigma$.

\end{document}